\documentclass[twocolumn,pra,superscriptaddress]{revtex4}
\usepackage[dvipsnames,usenames]{color}
\usepackage{graphicx}
\usepackage{dcolumn}
\usepackage{bm}
\usepackage{amssymb}
\usepackage{amsmath}
\usepackage{extarrows}
\usepackage{color}
\begin{document}

\title{Topological metal bands with double-triple-point fermions in optical lattices}

\author{Xue-Ying Mai}
\affiliation{Guangdong Provincial Key Laboratory of Quantum Engineering and Quantum Materials,
SPTE, South China Normal University, Guangzhou 510006, China}

\author{Yan-Qing Zhu}
\affiliation{National Laboratory of Solid
State Microstructures and School of Physics, Nanjing University,
Nanjing 210093, China}

\author{Zhi Li}
\affiliation{Guangdong Provincial Key Laboratory of Quantum Engineering and Quantum Materials,
SPTE, South China Normal University, Guangzhou 510006, China}

\author{Dan-Wei Zhang}
\email{danweizhang@m.scnu.edu.cn}\affiliation{Guangdong Provincial Key Laboratory of Quantum Engineering and Quantum Materials,
SPTE, South China Normal University, Guangzhou 510006, China}

\author{Shi-Liang Zhu}
\email{slzhu@nju.edu.cn} \affiliation{National Laboratory of Solid
State Microstructures and School of Physics, Nanjing University,
Nanjing 210093, China} \affiliation{Guangdong Provincial Key Laboratory of Quantum Engineering and Quantum Materials,
SPTE, South China Normal University, Guangzhou 510006, China}
%\affiliation{Synergetic Innovation Center of Quantum Information and Quantum
%Physics, University of Science and Technology of China, Hefei
%230026, China}

\begin{abstract}
Novel fermionic quasiparticles with integer pseudospins in some
energy bands, such as pseudospin-1 triple-point fermions, recently
attract increasing interest since they are beyond the conventional
spin-$1/2$ Dirac and Weyl counterparts. In this paper, we propose
a class of pseudospin-1 fermioic excitations emerging in
topological metal bands, dubbed double-triple-point (DTP)
fermions. We first present a general three-band continuum model
with $C_4$ symmetry in three dimensions, which has three types of
threefold degenerate points in the bands classified by their
topological charges $C=\pm4,\pm2,0$, respectively. They are dubbed
DTPs as spin-1 generalization of double-Weyl points. We then
construct two-dimensional and three-dimensional tight-binding
lattice models of topological metal bands with exotic DTP fermions
near the DTPs. In two dimensions, the band gaps close at a trivial DTP
with zero Berry phase, which occurs at the transition between the
normal and topological insulator phases. In three dimensions, the
topological properties of three different DTP fermions in lattice
systems are further investigated, and the effects of breaking
$C_4$ symmetry are also studied, which generally leads to
splitting each quadratic DTP into two linear triple points and
gives topological phase diagrams. Using ultracold fermionic atoms
in optical lattices, the proposed models can be realized and the
topological properties of the DTP fermions can be detected.
\end{abstract}

\date{\today}

\maketitle

\section{introduction}
\begin{figure}[tbph]
\centering
\includegraphics[width=6cm]{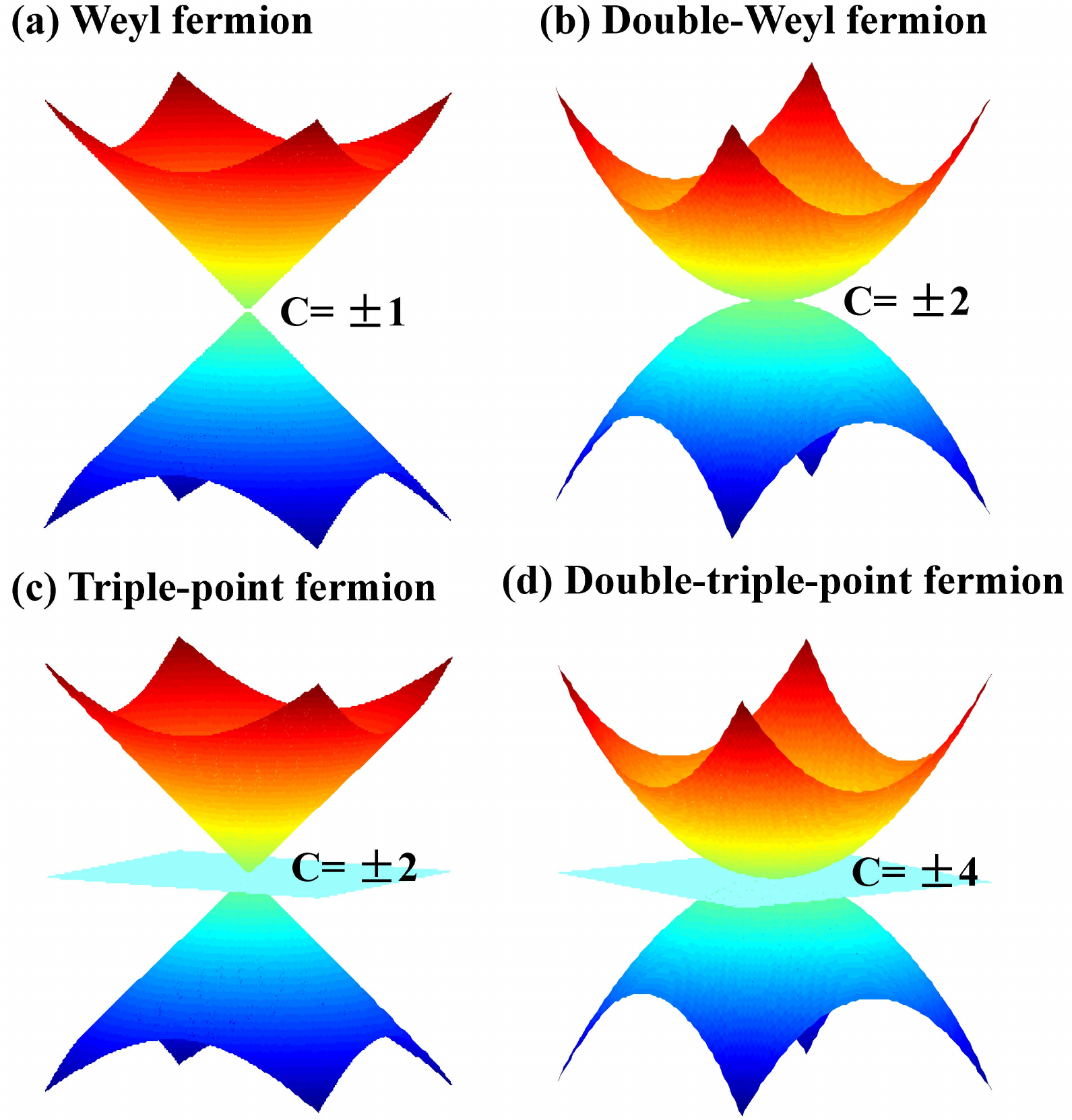}
\caption{(Color online) Energy dispersions of  (a) Spin-1/2 Weyl
fermions; (b) Spin-1/2 double-Weyl fermions; (c) Spin-1
triple-point fermions; (d) Spin-1 double-triple-point fermions.
The corresponding topological charges of the band degenerate
points are shown.} \label{four-dispersion}
\end{figure}

Topological semimetals and metals, such as Dirac and Weyl semimetals, have recently attracted broad attentions due to their interesting fundamental physics and potential applications \cite{Armitage}. In three-dimensional (3D) Weyl semimetals, the low-energy excitations near twofold band crossings dubbed as Weyl points resemble the well-known Weyl fermions in particle physics \cite{WeylTheo1,Burkov1,Balents,Burkov2,Zyuzin,Parameswaran,Riemann,SMExp1,SMExp2}. The Weyl fermions have linear dispersion along all three momenta directions [see Fig. \ref{four-dispersion}(a)] governed by the effective Weyl Hamiltonian, and a pair of Weyl points carry topological charges $C=\pm1$, which support gapless Fermi arc surface states. Both the Weyl fermions (points) in the bulk and Fermi arcs in the surface give rise to exotic phenomena, such as anomalous electromagnetic responses \cite{Armitage}. In another class of Weyl semimetals, the multi-Weyl semimetals \cite{Xu2011,Fang2012,Huang2016}, the twofold band degeneracies carry topological charges of higher magnitudes and can be stabilized by certain point-group crystal symmetries. Particularly, the double-Weyl points in double-Weyl semimetals have topological charges $C=\pm2$ and linear dispersion only along one dimension but quadratic along the other two dimensions [see Fig. \ref{four-dispersion}(b)], near which the low-energy excitations are called as double-Weyl fermions \cite{Xu2011,Fang2012,Huang2016}.

Very recently,  considerable attention have been paid to searching
for unconventional massless fermionic excitations beyond Dirac and
Weyl paradigm
\cite{Bradlyn2016,Lv2016,Wieder,Weng,Winkler,Bel,Tang,Minimal,Xu2017,Ma2018,Maxwell,Fallani,HPHu2017,Lan2011,Liang2016}.
In contrast to particles in high-energy physics constrained by
Poincar\'{e} symmetry, quasiparticles in a lattice system are only
constrained by certain subgroups (space groups) of the
Poincar\'{e} symmetry, which allows the emergence of ``new
fermions" (fermionic quasiparticles beyond the
Dirac-Weyl-Majorana classification) in some band structures with
three- or morefold degeneracies \cite{Bradlyn2016}.%, which code the (pseudo)spin degrees of freedom.
In particular, triple-point (three-component) fermions as spin-1
generalization of Weyl fermions in certain topological metal bands
with threefold degeneracies were theoretically predicted and then
experimental observed in some condensed matter materials
\cite{Lv2016,Ma2018}. Similar to the spin-1/2 Weyl fermions, the
spin-1 triple-point fermions have linear dispersion along all
momentum directions near the threefold degenerate points, which
carry topological charges $C=\pm2$ \cite{Bradlyn2016,Maxwell} [see
Fig. \ref{four-dispersion}(c)]. Along this direction, an important
question then is whether other types of triple-point fermions can
emerge in topological metal bands, such as the spin-1
generalization of double-Weyl fermions [see Fig.
\ref{four-dispersion}(d)] that are yet to be studied.

On the other hand, it has been shown that ultracold atoms in optical lattices with synthetic gauge fields and spin-orbit coupling provide a promising platform for studying topological states and phenomena \cite{GaugeRMP,GaugeRPP,SOC-Review1,SOC-Review2,Cooper2018,ZDW2018}. Remarkably, Dirac fermions in a tunable honeycomb optical lattice \cite{Zhu2007,Lim} has been realized \cite{Tarruell2012}, where the topological phase transition and the $\pi$ Berry phase of a Dirac point have been directly observed \cite{Duca2015}. The celebrated Harper-Hofstadter model \cite{HHModel} and Haldane model \cite{HaldaneModel} have been realized with cold atoms in two-dimensional (2D) optical lattices \cite{Miyake,Bloch2013a,Bloch2015,Shao,Jotzu}, where the Chern numbers characterizing the band topology have also been measured. The experimental observation of chiral edge states with cold atoms in synthetic Hall ribbons has been reported \cite{Cold-Edge1,Cold-Edge2}. The topological (geometric) pumping \cite{Thouless,Wang} has been demonstrated in optical superlattices \cite{Pumping1,Pumping2,Pumping3,Pumping4}. The 2D spin-orbit coupling and topological bands have been generated in an optical Raman lattice \cite{Liu2014,2DSOC}. In addition, several schemes have been suggested to realize topological semimetal bands with Weyl fermions \cite{Jiang,Xu2014,Liu2015,Dubcek,ZDW2015,Ganeshan,He,ZLi,Xu1,WeylII1,Shastri} and double-Weyl fermions \cite{Lepori,My}. Moreover, it has been proposed that the spin-1 triple-point fermions can emerge in some topological metal bands in cold atom systems \cite{Maxwell,Fallani,HPHu2017}, which can even be simulated in parameter space \cite{Tan2018,HPHu2018}. Due to the elusive nature of triple-point fermions in real materials and their exotic properties \cite{Bradlyn2016,Lv2016,Wieder,Weng,Winkler,Bel,Tang,Minimal,Xu2017,Ma2018,Maxwell,Fallani,HPHu2017}, proposals for realizing other types of triple-point fermions in artifical cold atom systems would be of great value.

In this paper, we propose a new class of spin-1 fermioic excitations emerging in topological metal bands, which are dubbed double-triple-point (DTP) fermions as the spin-1 generalization of double-Weyl fermions. We first present a general three-band continuum model with $C_4$ symmetry in 3D, which is shown to contain three different types of threefold degenerate points in the bands classified by their topological charges $C=\pm4,\pm2,0$, respectively. The threefold degenerate points are called as DTPs as spin-1 double-Weyl points. We then construct 2D and 3D tight-binding lattice models of topological metal bands with DTP fermions near the DTPs, respectively. In 2D, the bands close at a trivial DTP with zero Berry phase, which occurs at the transition between the normal and topological insulator phases. In 3D, the topological properties of three different DTP fermions in lattice systems are further investigated, such as the bulk-edge correspondence. The effects of breaking $C_4$ symmetry in this case are also studied, which generally leads to splitting each DTP into two triple points and gives topological phase diagrams. Finally, we discuss the realization of the models and detection of the topological properties of the DTP fermions with ultracold fermionic atoms in optical lattices.

The paper is organized as follows. In Sec. II, we construct a continuum model with three types of DTPs. In Section III, we study DTP fermions in a 2D tight-binding lattice model. In Sec. IV, the topological properties of three different DTP fermions in 3D lattice systems are further investigated. In Sec. V, we propose realization of the models and detection of the the DTP fermions in optical lattices. A brief conclusion is finally given in Sec. VI.

\section{A continuum model}
Let us begin with a three-band continuum model with a symmetry $C_4$ in 3D, which takes the following Hamiltonian
\begin{equation}\label{DTPHam}
\mathcal{H}(\boldsymbol{k})=(k_y^2-k_x^2)S_x+k_xk_yS_y+k_z(\alpha S_z+\beta N_{ij}),
\end{equation}
where $\alpha$ and $\beta$ are parameters, $\mathbf{S}=(S_x,S_y,S_z)$ are the spin-1 matrices given by \cite{Maxwell}
\begin{equation}
\begin{split}\label{spin}
S_x=
\left(\begin{matrix}
0&0&0\\
0&0&-i\\
0&i&0
\end{matrix}
\right),  S_y=
\left(\begin{matrix}
    0&0&i\\
    0&0&0\\
    -i&0&0
    \end{matrix}
\right),  S_z=
\left(\begin{matrix}
    0&-i&0\\
    i&0&0\\
    0&0&0
    \end{matrix}
\right)
\end{split},
\end{equation}
and $N_{ij}$ are the tensors given by \cite{HPHu2017} $N_{ij}=(S_iS_j+S_jS_i)/2-\delta_{ij}\mathbf{S}^2/3$ $(i,j=x,y,z)$.
Note that the spin-1 matrices satisfy that $[S_x,S_y]=iS_z$,
$\mathbf{S}\times\mathbf{S}=i\mathbf{S}$, and $\mathbf{S}^2=S_x^2+S_y^2+S_z^2=S(S+1)$ with $S=1$. This model preserves $C_{4}$ symmetry represented as
\begin{eqnarray}
C_4\mathcal{H}(\boldsymbol{k})C_{4}^{-1}=\mathcal{H}(R\boldsymbol{k}),
\end{eqnarray}
where $C_4=e^{-i\pi S_z}$ is a point-group operator for the fourfold rotation about $z$ axis and $R$ is an operator taking $(k_x,k_y,k_z)$ to $(k_y,-k_x,k_z)$. In addition, as $T\mathcal{H}(\boldsymbol{k})T^{-1}\neq\mathcal{H}(-\boldsymbol{k})$, where the time-reversal operator $T=\hat{I}\hat{K}$ with
$\hat{I}=\text{diag}(1,1,1)$ and $\hat{K}$ being the complex conjugate operator, the model does not have time reversal symmetry.

One can find that $\mathcal{H}(\boldsymbol{k})$  exhibits a
threefold degenerate point at $\boldsymbol{k}=0$. As shown in Fig.
\ref{four-dispersion}(d) with $\beta=0$, near the degenerate point
of the three bands, the dispersions along the $k_x$ and $k_y$ are
quadratic while linear along $k_z$, which is a triple-point with
analogous quadratic dispersion of a double Weyl-point. Thus we
name such a new triple-point as DTP, which carries a topological
charge $C$ that can be defined in terms of the first Chern number
on a surface enclosing the point:
\begin{equation}\label{ChN}
C_n=\frac{1}{2\pi}\oint_{\mathcal{S}}\mathbf{F}_n(\boldsymbol{k})\cdot d\mathcal{S}.
\end{equation}
Here $\mathcal{S}$ denotes the integration surface,  and
$\mathbf{F}_n(\boldsymbol{k})=\nabla\times\langle{u_n(\boldsymbol{k})|i\partial_k|u_n(\boldsymbol{k})\rangle}$
is the Berry curvature of the $n$-th band with the wave function
$|u_n(\boldsymbol{k})\rangle$, where the band indices $n$ for the
lower, middle, and higher bands are denoted as $-$, $0$, and $+$,
respectively. For $\beta=0$, one has the simple expression
$\mathbf{F}_n(\boldsymbol{k})=-2n\text{sign}(\alpha)\boldsymbol{k}/|\boldsymbol{k}|^3$
and $C_n=-4n\text{sign}(\alpha)$ for the $n$-th band. Hereafter we
use the lower-band Chern number to label the DTPs: $C=C_-$. The
simplest DTP in this case with topological charges $C=\pm4$ is
called as type-I DTP. Note that $C_+=-C_-$ for the higher and
lower bands and $C_0=0$ for the middle one even if $\beta N_{ij}$
term exists, which has been numerical confirmed.

Similar to the case for the linear triple-points \cite{HPHu2017},
we find that there are three types of quadratic DTPs induced by
the tensors $N_{ij}$ of different forms in Eq. (\ref{DTPHam}).
First, the monopole charges of a type-I DTP with $C=\pm4$ will not
change with the three spin-tensors $N_{xx}$, $N_{yy}$, and
$N_{xy}$. Second, when $|\beta|>|\alpha|\neq0$, the tensor
$N_{zz}$ induces a DTP with $C=\pm2$ which is named type-II DTP.
Finally, when $|\beta|>2|\alpha|\neq0$, a trivial type-III DTP
with $C=0$ can be induced by the tensor $N_{xz}$ or $N_{yz}$.

\begin{figure}[tbph]
\centering
\includegraphics[width=8cm]{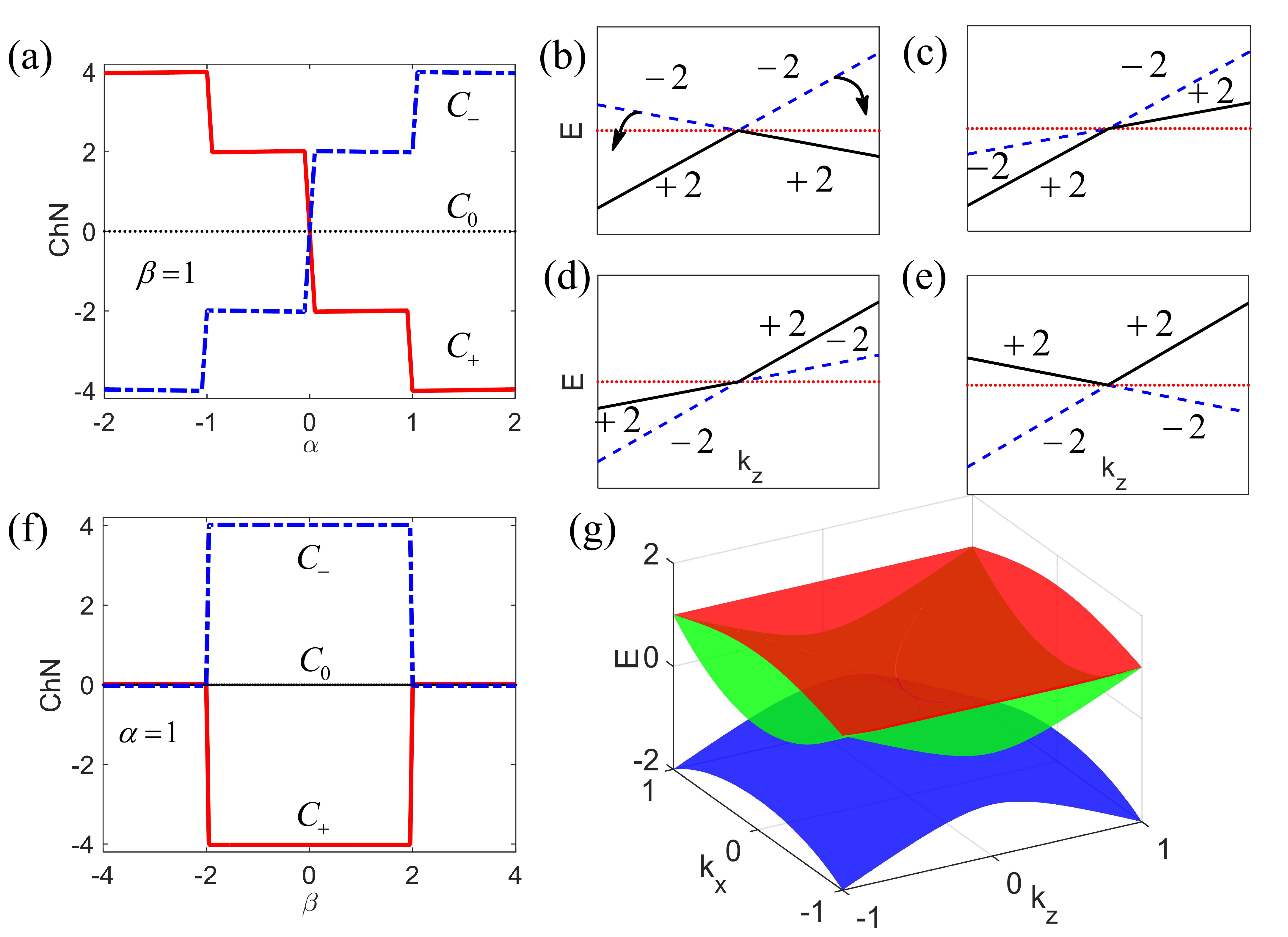}
\caption{(Color online). Phase transitions between type-I and
type-II (type-III) DTPs by tuning $\alpha$ ($\beta$) while fixing
$\beta=1$ ($\alpha=1$) in Eq. (\ref{DTPHam}). Chern numbers as
functions of $\alpha$ ($\beta$) for the lowest (dashed blue),
middle (dotted black), and higher (solid red) bands are shown in
(a) and (f), respectively. (b)-(e) Band structures along the
$k_x=k_y=0$ line with $\alpha = 2$, $0.5$, $-0.5$, and $-2$,
respectively. The Chern number contributions for each branch are
$+2$ (solid black), $-2$ (dashed blue), and $0$ (dotted red). (g)
Bands crossing at one of the two transition points with $\beta=2$,
where the lowest, middle, higher bands are labeled by the color
blue, green, and red, respectively. The levels crossing only
between the middle and higher bands.} \label{phasetransition}
\end{figure}

Figure \ref{phasetransition}(a) shows a phase transition between
type-I and type-II DTPs which can be induced through the
competition between parameters $\alpha$ and $\beta$ with the
spin-tensor $N_{ij}=N_{zz}$ in Eq. (\ref{DTPHam}). In this case,
the lowest-band Chern number changes from $4$ (type-I) to $\pm2$
(type-II), and then to $-4$ (type-I) with decreasing $\alpha$ by
fixing $\beta=1$. As depicted in Figs.
\ref{phasetransition}(b)-(e), the band structure along the
$k_x=k_y=0$ line is concerned. Notably, the Chern number of each
band has two contributions from the $k_z>0$ and $k_z<0$ branches.
For $\alpha>1$, the $k_zS_z$ term dominates, and the band touching
point is still the type-I DTP with $C=4$ because of the Chern
number contributions from the two branches of the lowest band are
both $+2$ [Fig. \ref{phasetransition}(b)]. Decreasing $\alpha$,
the $k_z>0$ ($k_z<0$) branch of the lowest band rotates
counterclockwise (clockwise).  At the critical value $\alpha=1$,
the middle band simultaneously crosses the $k_z>0$ branch of the
lowest band and $k_z<0$ branch of the higher band. After that, the
two branches of lowest band contribute $C_{k_z>0}=0$ and
$C_{k_z<0}=2$, as sketched in Fig. \ref{phasetransition}(c), give
rise to a type-II DTP with $C=2$, which is consistent with the
numerical results in Fig. \ref{phasetransition}(a). If one further
decreases $\alpha$ to $0$, another band crossing will occur
between the middle band and the $k_z>0$ ($k_z<0$) branch of higher
(lowest) band [Fig. \ref{phasetransition}(d)]. However, the DTP is
still type-II since this process only changes the sign of $C$ to
$-2$. The last transition point with level crossing occurs at
$\alpha=-1$. Compared with the case $\alpha=1$, all bands are
simply reversed when $\alpha<-1$ as depicted in Fig.
\ref{phasetransition}(e), and the DTP with $C=-4$ belongs to
type-I.

%%%%%%%%%%%%%%%%%%%%%%
%%%%%%%%%%%%%%%%%%%%%%

\begin{figure}[tbph]
\centering
\includegraphics[width=8cm]{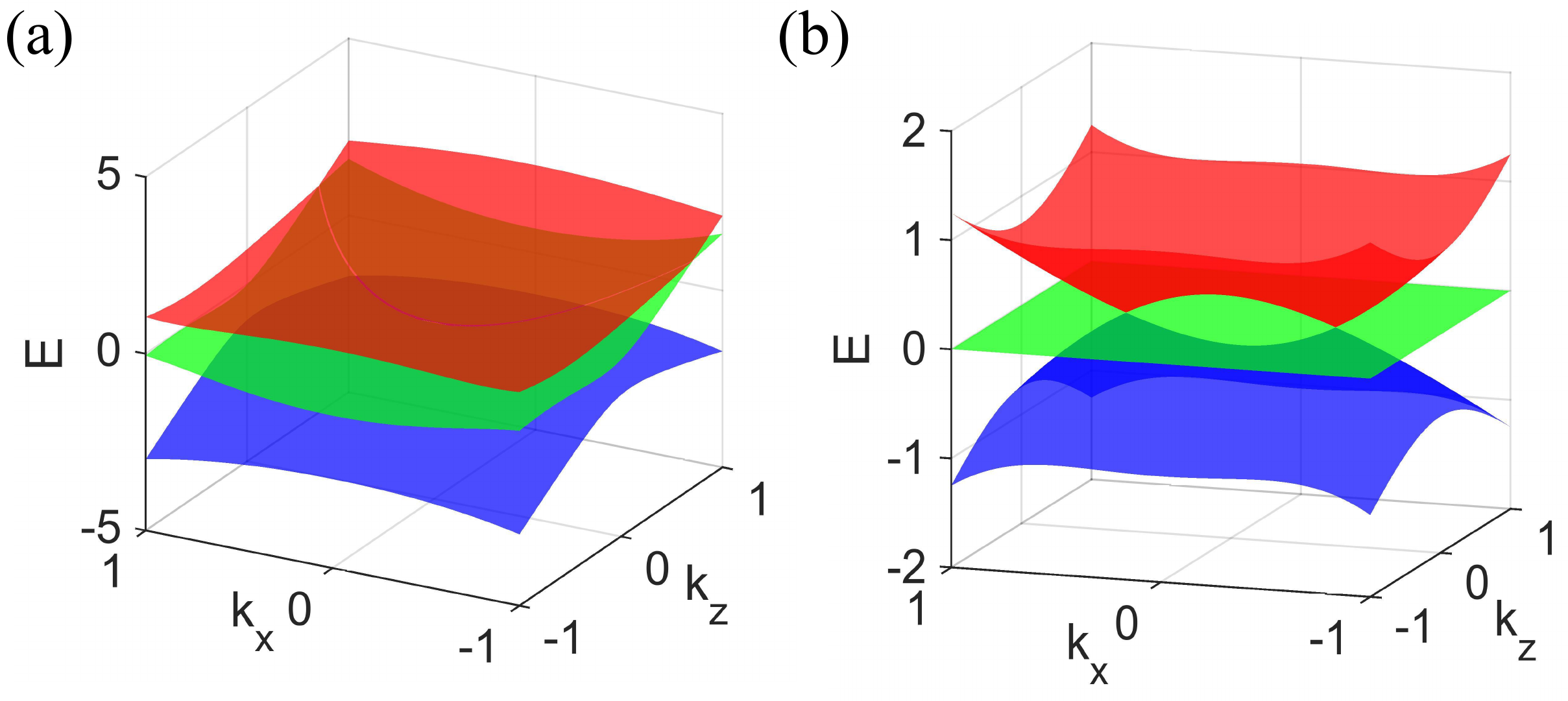}
\caption{(Color online). (a) Bands  crossing at the transition
point for $\alpha=1$ and $\beta N_{ij}=c_1N_{zz}+c_2N_{xz}$ in the
Hamiltonian (\ref{DTPHam}) with $c_1=1$ and $c_2=2\sqrt{2}$. The
levels crossing only between the middle and highest bands along one
quadratic curve at the $k_z>0$ branch. (b) For $\alpha=1$ and
$\beta=0$, a quadratic type-I DTP ($C=\pm{4}$) splits into two
linear TPs carrying the same monopole ($C=\pm2$).}
\label{TypeII-III}
\end{figure}

The spin-tensors $N_{xz}$ or $N_{yz}$ can  induce a type-III DTP
with $C=0$ from a type-I DTP. To illustrate such a phase
transition, we choose $N_{ij}=N_{xz}$ with $\alpha=1$ in
Hamiltonian (\ref{DTPHam}). The DTP is of type-I for $|\beta|<2$
and type-III for $|\beta|>2$, respectively. At $\beta=2$, as
illustrated in Fig. \ref{phasetransition}(g), the higher and
middle bands cross along two quadratic curves $k_z
{\pm}k_x^2=k_y=0$, while the lowest band touches with these two
bands at $\boldsymbol{k}=0$. Near one of these two curve nodes,
e.g., $k_z-k_x^2=k_y=0$, the spectra are found to be
$\beta{k_z}/2$ and $(-\beta\pm\sqrt{\beta^2+32})k_z/4$. The level
crossing in the vicinity of the other curve node is similar.
Moreover, for the case of $\beta=-2$, the lowest and middle bands
cross along the two curves and the highest band touches with these
two bands at $\boldsymbol{k}=0$. Because each band crossing
changes the Chern number by $2$, and the crossings along the two
curves are in the same branch, the Chern number must change by
$4$. Such a topological transition of a DTP from type-I to
type-III is shown in Fig. \ref{phasetransition}(f). In contrast,
the Chern number in the transition from type-II to type-III DTPs
only changes by $2$, and the band crossing is along only one
curve. For instance, we consider $\alpha=1$ and $\beta
N_{ij}=c_1N_{zz}+c_2N_{xz}$ in the Hamiltonian (\ref{DTPHam}) with
$c_{1,2}> 0$, the transition occurs at $c_1=c_2^2/4-1$, and the
band crossing is along the $k_x^2-c_2k_z/2=k_y=0$ curve, as shown
in Fig. \ref{TypeII-III}(a).

By introducing a $C_4$-breaking term,  a quadratic DTP of monopole
charge $C$ can be split into two linear triple-points with
monopole charge $C/2$. For concreteness, let us consider adding a
perturbation term $\lambda^2S_x$ in the model Hamiltonian
(\ref{DTPHam}), then the bands touch at two threefold degenerate
points $\boldsymbol{K}_{\pm}=(\pm|\lambda|,0,0)$. Near these two
points, the Hamiltonian can be expanded to the linear order with
$\boldsymbol{q}=\boldsymbol{k}-\boldsymbol{K}_\pm$ as
\begin{equation}\label{TPHam}
\mathcal{H}_{\pm}(\mathbf{\boldsymbol{q}})=\mp2|\lambda|q_xS_x\pm|\lambda|q_yS_y+q_z(\alpha{S_z}+\beta{N}_{ij}).
\end{equation}
A threefold degenerate point described by this effective Hamiltonian is a linear triple point \cite{HPHu2017}. Specially, the type-I DTP of $C=\pm4$ for $N_{ij}=N_{zz}$ are broken into two linear triple point with $\mathcal{C}=\pm2$, as shown in Fig. \ref{TypeII-III}(b). The effects of $C_4$-breaking terms for type-II and type-III are similar.

%%%%%%%%%%%%%%%%%%%%%%
%%%%%%%%%%%%%%%%%%%%%%

\section{DTP fermions in 2D lattice model}

\begin{figure}[tbph]
\centering
\includegraphics[width=8cm]{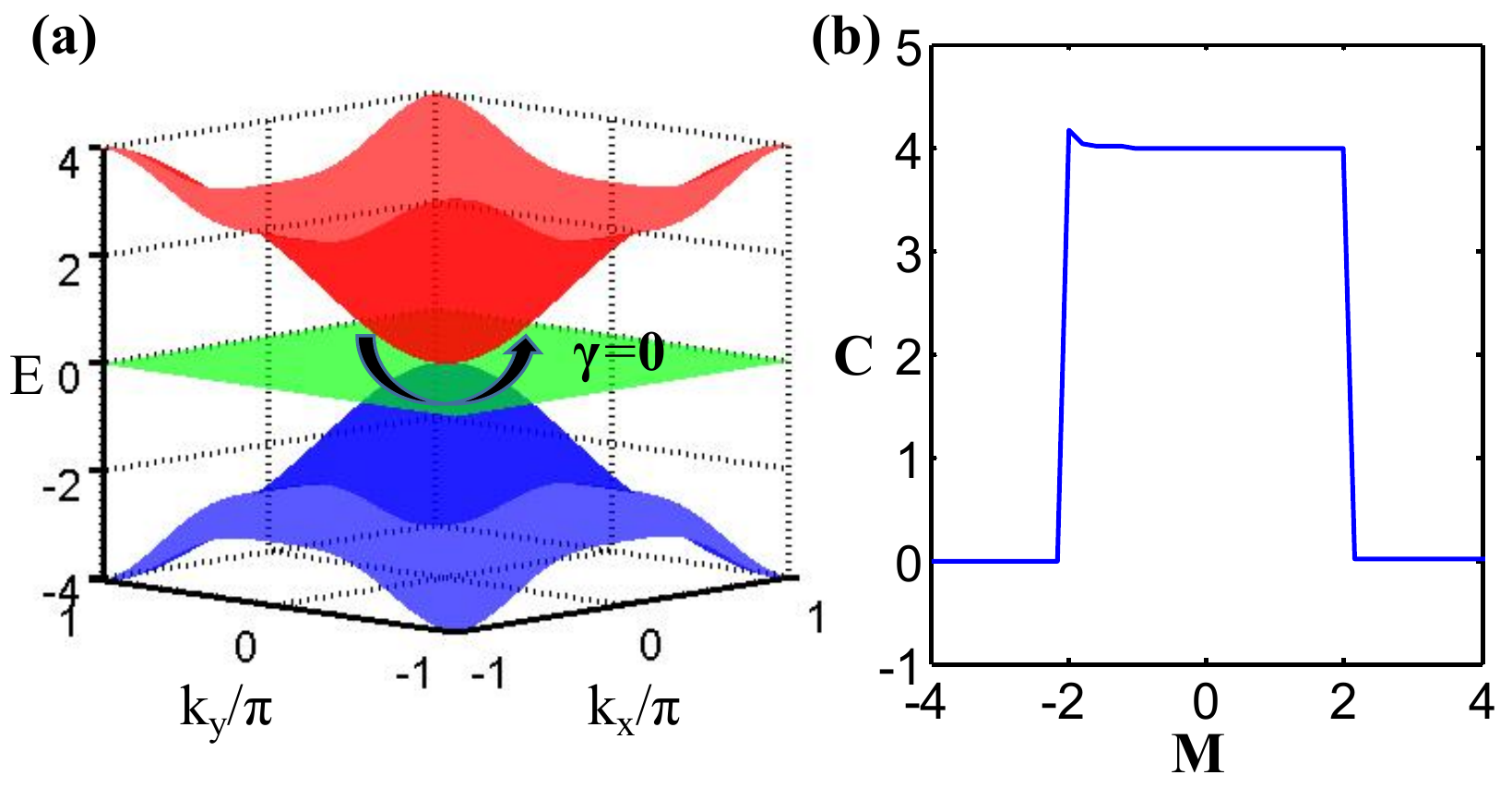}
\caption{(Color online) (a) The band dispersion in the
$k_{x}-k_{y}$ plane with $M=2$. (b) The Chern numbers as a
function of $M$. When $M\neq\pm2$, the 2D system is in the
topological or trivial insulating phase.} \label{band}
\end{figure}

For a lattice system that is filled with fermionic particles and
effectively described by the low-energy Hamiltonian
(\ref{DTPHam}), the quasiparticle excitations near the DTPs can be
named DTP fermions with (pseudo)spin-1 and quadratic dispersion.
In this section, we first study DTP fermions in a 2D lattice
model, which can reduce into the Hamiltonian (\ref{DTPHam})
without $k_z$ terms. The three spin-1 states are labeled as
$|\uparrow\rangle$, $|0\rangle$, $|\downarrow\rangle$, and the 2D
tight-binding Hamiltonian on a square lattice is given by
\begin{eqnarray} \label{2DHam}
\nonumber\hat{H}_{2D}&=&t\sum_{\mathbf{r}}\left[
\hat{H}_{\mathbf{rx}}+\hat{H}_{\mathbf{ry}}+\hat{H}_{\mathbf{rxy}}+\hat{H}_{\mathbf{M}}
 \right],\\
\nonumber\hat{H}_{\mathbf{rx}}&=&\left[-\frac{i}{2}(\hat{a}^{\dag}_{\mathbf{r},0}\hat{a}_{\mathbf{r+x},
\uparrow}+\hat{a}^{\dag}_{\mathbf{r},0}\hat{a}_{\mathbf{r-x},
\uparrow})+\textrm{H.c.} \right]\\
\nonumber&&+\left[\frac{i}{2}(\hat{a}^{\dag}_{\mathbf{r},\downarrow}\hat{a}_{\mathbf{r+x},
0}+\hat{a}^{\dag}_{\mathbf{r},\downarrow}\hat{a}_{\mathbf{r-x},
0})+\textrm{H.c.} \right],\\
\hat{H}_{\mathbf{ry}}&=&\left[-\frac{i}{2}(\hat{a}^{\dag}_{\mathbf{r},0}\hat{a}_{\mathbf{r+y},
\uparrow}+\hat{a}^{\dag}_{\mathbf{r},0}\hat{a}_{\mathbf{r-y},
\uparrow})+\textrm{H.c.} \right]\\
\nonumber&&+\left[\frac{i}{2}(\hat{a}^{\dag}_{\mathbf{r},\downarrow}\hat{a}_{\mathbf{r+y},
0}+\hat{a}^{\dag}_{\mathbf{r},\downarrow}\hat{a}_{\mathbf{r-y},
0})+\textrm{H.c.} \right],\\
\nonumber\hat{H}_{\mathbf{rxy}}&=&\left[-\frac{i}{4}(\hat{a}^{\dag}_{\mathbf{r},\downarrow}\hat{a}_{\mathbf{r+(x-y)},
\uparrow}+\hat{a}^{\dag}_{\mathbf{r},\downarrow}\hat{a}_{\mathbf{r-(x-y)},
\uparrow})+\textrm{H.c.} \right]\\
\nonumber&&+\left[\frac{i}{4}(\hat{a}^{\dag}_{\mathbf{r},\downarrow}\hat{a}_{\mathbf{r+(x+y)},
\uparrow}+\hat{a}^{\dag}_{\mathbf{r},\downarrow}\hat{a}_{\mathbf{r-(x+y)},
\uparrow})+\textrm{H.c.} \right],\\
\nonumber\hat{H}_{\mathbf{M}}&=&iM\hat{a}^{\dag}_{\mathbf{r},0}\hat{a}_{\mathbf{r},
\uparrow}+\textrm{H.c.},
\end{eqnarray}
where $\hat{a}_{\mathbf{r},s}$($\hat{a}^{\dag}_{\mathbf{r},s}$) is the annihilation (creation) operator on site $\mathbf{r}$ with spin $s=\{\uparrow,0,\downarrow\}$, $M$ is the strength of an effective Zeeman potential, and the hopping strength is set $t=1$ as the energy unit hereafter. The terms $\hat{H}_{\mathbf{rx}}$, $\hat{H}_{\mathbf{ry}}$, and $\hat{H}_{\mathbf{rxy}}$ represent the spin-flip hopping along $x$, $y$, and $x\pm y$ axis, respectively. Under the periodic boundary condition, the tight-binding Hamiltonian can be written as $\hat{H}_{2D}=\sum_{\mathbf{k},ss'}\hat{a}^{\dag}_{\mathbf{k}s}[\mathcal{H}(\mathbf{k})]_{ss'}\hat{a}_{\mathbf{k}s'}$,
where $\hat{a}_{\boldsymbol{k}s}=1/\sqrt{V}\sum_{\boldsymbol{r}}e^{-i\boldsymbol{k\cdot
r}}\hat{a}_{\boldsymbol{r}s}$ is the annihilation operator in momentum space $\boldsymbol{k}=(k_x,k_y)$, and the Bloch Hamiltonian is written as (the lattice spacing $a\equiv1$ and $\hbar\equiv1$)
\begin{eqnarray}
\mathcal{H}(\boldsymbol{k})&=&\mathbf{d}(\boldsymbol{k})\cdot \mathbf{S}, \nonumber \\
d_x &=& \cos{k_x}-\cos{k_y},\nonumber \\
d_y &=& \sin{k_x}\sin{k_y},\\
d_z &=& M-\cos{k_x}-\cos{k_y}.\nonumber
\end{eqnarray}
The energy spectrum is given by $E(\boldsymbol{k})=0,\pm|\mathbf{d}(\boldsymbol{k})|$, which contains a zero-energy flat band in the middle of the three bands. For $M=2$ ($M=-2$), the three bands touch at a single point $\mathbf{K}_+=(0,0)$ [$\mathbf{K}_-=(\pi,\pi)$] in the first Brillouin zone, as shown in Fig. \ref{band}(a). Expanding the Bloch Hamiltonian near the threefold degenerate point with $\mathbf{q}=\mathbf{k}-\mathbf{K}_\pm$ yields the low-energy effective Hamiltonian
\begin{equation}
\mathcal{H}_{\text{eff}}^{\pm}(\mathbf{q})=\pm[\frac{1}{2}(q^{2}_y-q^{2}_x)S_x+q_xq_yS_y],
\end{equation}
indicating that this threefold degenerate point is a 2D DTP with
quadratic dispersion along $k_x$ and $k_y$ directions. When the
Fermi level lies near the 2D DTP, the lattice system is in a
metallic phase with emergent 2D DTP fermions.

\begin{figure}[tbph]
\centering
\includegraphics[width=8cm]{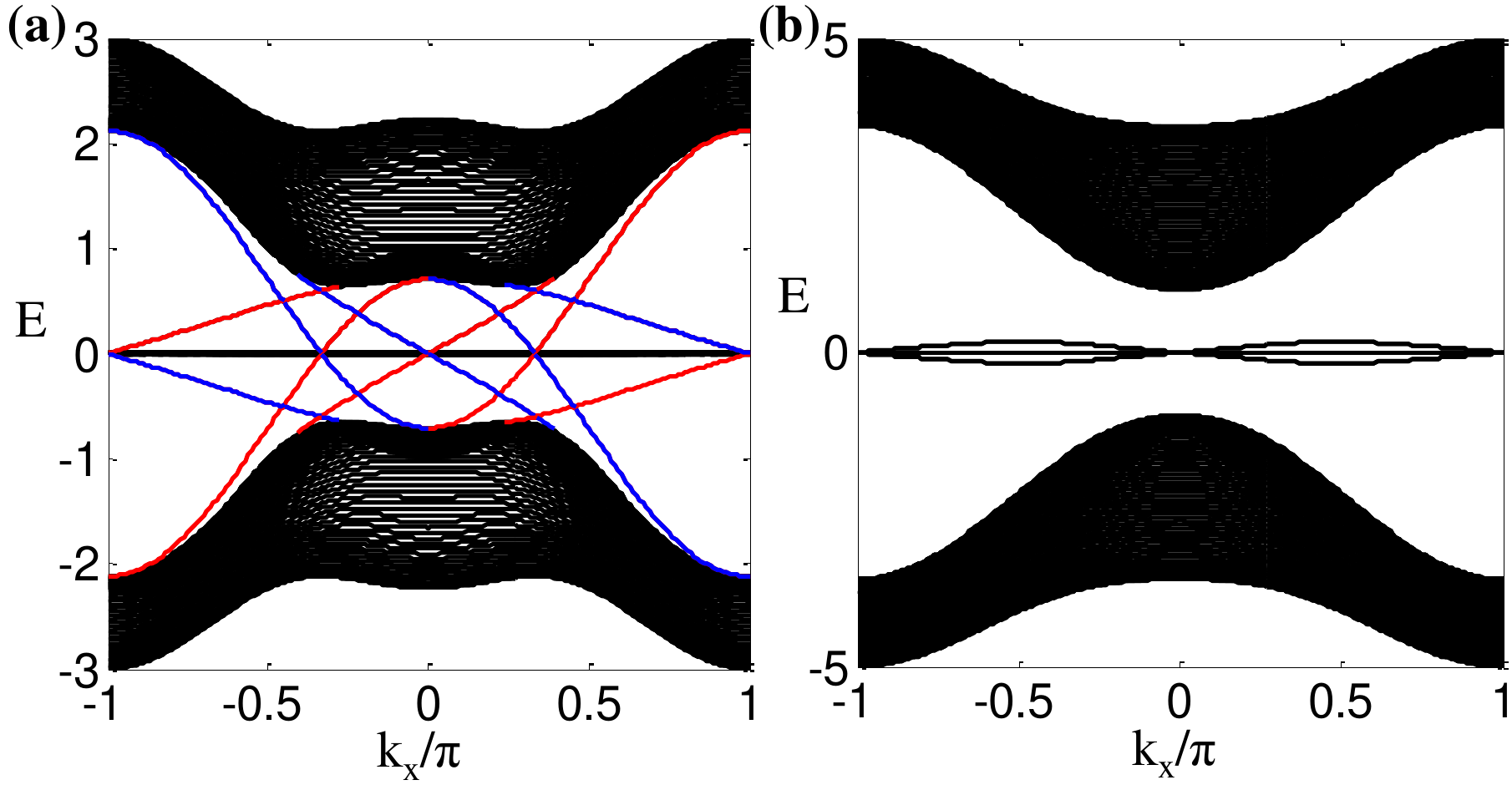}
\caption{(Color online) (a) The energy spectrum $E(k_x)$ of reduced chain with lattice sites $L_y=60$ under open boundaries for (a) $M=1$ and (b) $M=3$. The red and blue colors denote the edge modes in opposite edges.} \label{energy-spectrum}
\end{figure}

To further study the topological properties of the system, we first calculate the Berry phase of the 2D DTP point for $M=\pm2$ by circling around it
\begin{equation}
\begin{aligned}
\gamma=\oint_{C}d\boldsymbol{k}\cdot\mathbf{A(\boldsymbol{k})},
\end{aligned}
\end{equation}
where the Berry connection $\mathbf{A(\boldsymbol{k})}=i\langle u_{-}(\boldsymbol{k})|\nabla_{\boldsymbol{k}}|u_{-}(\boldsymbol{k})\rangle$. The analytical calculation gives the Berry phase $\gamma=0$ for both $M=\pm2$ which is confirmed by numerical integration. Thus, the 2D DTPs are topologically trivial, and the 2D DTP fermions in this model are not topological quasiparticles. When $M\neq\pm2$, the system changes from a trivial metallic state to an insulator state with band gaps. We then calculate the Chern number of the lowest gapped band
\begin{equation} \label{2DChN}
C=\frac{1}{2\pi}\int_{\text{BZ}}{dk_xdk_y}\frac{1}{d^{3}}\mathbf{d}\cdot(\partial_{k_x}{\mathbf{d}}\times\partial_{k_y}{\mathbf{d}}).
\end{equation}
The obtained $C$ as a function of the parameter $M$ is shown in Fig. \ref{band}(b), which indicates that the system is a topological Chern (trivial band) insulator with $C=4$ ($C=0$) when $|M|<2$ ($|M|>2$). We also numerically calculate the energy spectrum $E(k_x)$ under open boundary condition along $y$ direction with the lattice length $L_y=60$ for fixed $M=1$ and $M=3$, respectively. As shown in Figs. \ref{energy-spectrum}(a) and (b) for $M=1$ and $M=3$, there are four and zero chiral in-gap edge states connecting the three bulk bands, which is consistent of the bulk-edge correspondence for $C=4$ and $C=0$, respectively.

\section{DTP fermions in 3D lattice model}

We proceed to study the DTP fermions in a 3D cubic lattice system, which can reduce into the Hamiltonian (\ref{DTPHam}). The corresponding 3D tight-binding Hamiltonian is given by
\begin{equation} \label{3DHam}
\begin{split}
\hat{H}_{3D}&=~~\hat{H}_{2D}+\sum_{\mathbf{r}}\hat{H}_{\mathbf{rz}},\\
\hat{H}_{\mathbf{rz}}&=-\frac{i\alpha}{2}(\hat{a}^{\dag}_{\mathbf{r},0}\hat{a}_{\mathbf{r+z},\uparrow}+
\hat{a}^{\dag}_{\mathbf{r},0}\hat{a}_{\mathbf{r-z},\uparrow})+\textrm{H.c.},
\end{split}
\end{equation}
where $\hat{H}_{\mathbf{rz}}$ denotes the hopping term along the $z$ axis. The Bloch Hamiltonian of the 3D system for $\alpha=1$ takes the form as
\begin{eqnarray} \label{3DBlochHam}
\nonumber \mathcal{H}(\boldsymbol{k})&=&(\cos k_x-\cos k_y)S_x+\sin k_x\sin k_yS_y\\
&&+(M-\cos k_x-\cos k_y-\cos k_z)S_z.
\end{eqnarray}

\begin{figure}[tbph]
\centering
\includegraphics[width=8.7cm]{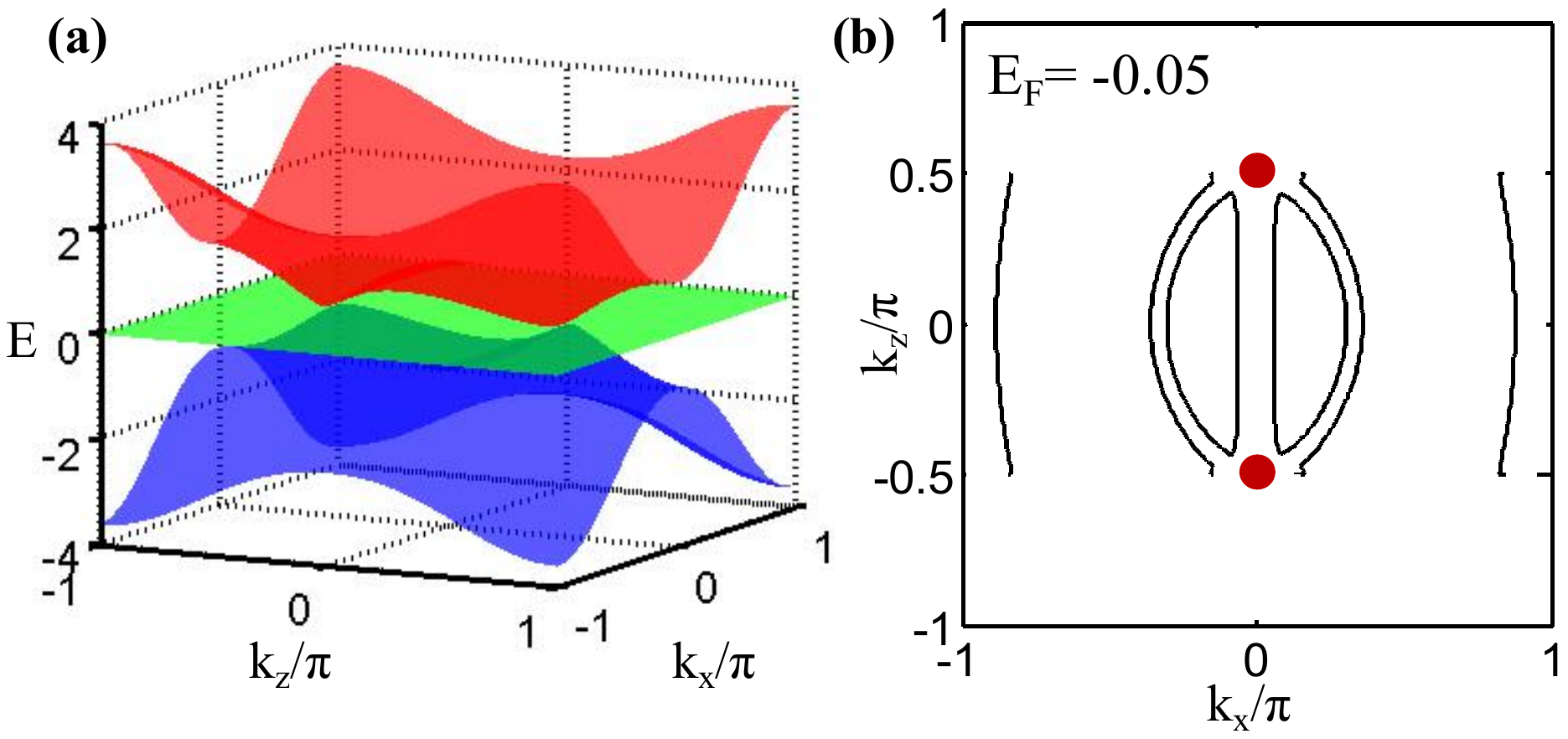}
\caption{(Color online) (a) The band dispersion as a function of $k_x$ and $k_z$ for $k_y=0$ and $M=2$. (b) The Fermi arcs appear at $E_{F}=-0.05$ under the open boundary condition along the $y$ direction. The black lines denote the surface states, and the two red dots denote the position of DTPs.} \label{fermi-arc}
\end{figure}

There are a pair of threefold degenerate points in the energy bands located at $\mathbf{T}_1^{\pm}=[0,0,\pm\arccos(M-2)]$ for $1<M<3$ as shown in Fig. \ref{fermi-arc}(a), and another pair at $\mathbf{T}_2^{\pm}=[\pi,\pi,\pm\arccos(M+2)]$ for $-3<M<-1$. One can obtain the
low-energy effective Hamiltonian around $\mathbf{T}_1^{\pm}$ with $\mathbf{q}=\mathbf{k}-\mathbf{T}_1^{\pm}$:
\begin{equation}
\mathcal{H}_{\text{eff}}^{\pm}(\mathbf{q})=\frac{1}{2}(q^{2}_y-q^{2}_x)S_x+q_xq_yS_y \pm v_zq_zS_z,
\end{equation}
where $v_z=\sqrt{1-(M-2)^{2}}$ for $1<M<3$. The monopole charge of the DTPs $\mathbf{T}_1^{\pm}$ can be obtained as $C=\pm4$, then they are type-I DTPs ($\mathbf{T}_2^{\pm}$ are also type-I DTPs). Since the 3D bulk bands are fully gapped when $k_z\neq\pm k^{c}_{z}$ with $k^{c}_{z}=\arccos(M-2)$, considering $k_z$ as a good quantum number, the dimension reduction Hamiltonian $\mathcal{H}_{k_z}(k_x,k_y)$ for a fixed $k_z\neq\pm k^{c}_{z}$ can be viewed as a 2D Chern insulator, which is topologically characterized by the $k_z$-dependent Chern number defined on the $k_x$-$k_y$ plane [see Eq. (\ref{2DChN})]:
\begin{eqnarray}\label{typeII1}
&&C_{k_z}=\left\{
                                  \begin{array}{ll}
                                    0, ~~& |k_z|>k^{c}_{z};\\
                                    4, ~~& 0<|k_z|<k^{c}_{z}.
                                  \end{array}
                                \right.
\end{eqnarray}
Thus the type-I DTPs with monopole charge $C=\pm4$ act as the topological transition points between two layer topological insulators with Chern number difference $\Delta C_{k_z}=\pm4=C$. Using open boundary condition along $y$ direction, we find that there are four Fermi arcs surface states effectively connecting the pair of DTPs, as shown in Fig. \ref{fermi-arc}(b). In these cases, the system has the topological metal bands with emergent DTP fermions.
At the critical points $M=\pm1$ or $\pm3$, the two DTPs merge and then disappear by opening a gap when $|M|<1$ and $|M|>3$. For $|M|<1$, the system is in a weak topological insulator phase, which has the Chern number $C_{k_z}=4$ for all $k_z$ and chiral surface states under open boundary condition. For $|M|>3$, it is a trivial insulating phase.

\begin{figure}[tbph]
\centering
\includegraphics[width=6.1cm]{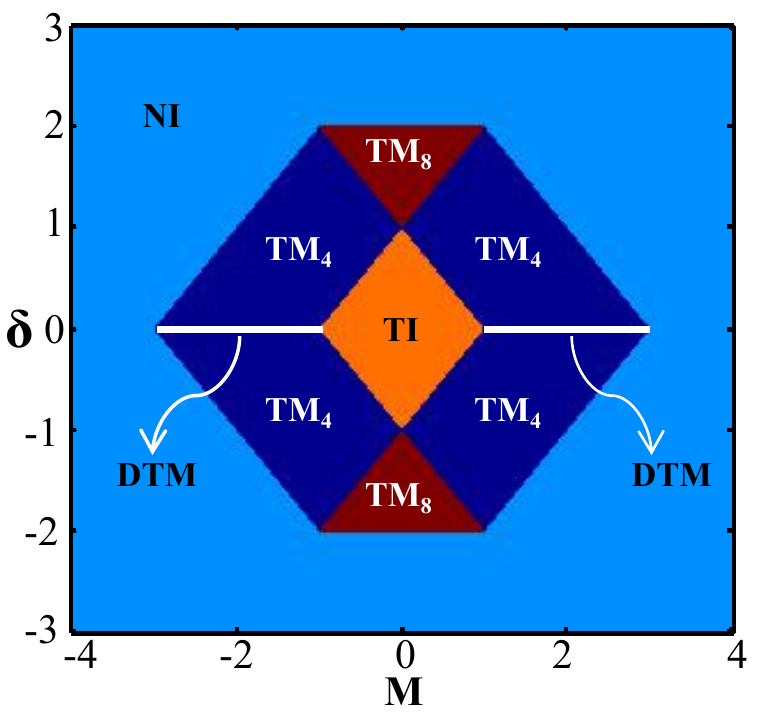}
\caption{(Color online) The phase diagram of the 3D lattice system with a $C_4$-symmetry breaking term. DTM denotes the DTP metal phase (white lines), TI (orange) and NI (light blue) denote a weak topological and normal insulating phase, TM$_{4}$ (dark blue) and TM$_{8}$ (red) represent the triple-point metal phase with four and eight triple points, respectively.} \label{phase-diagram}
\end{figure}

To further study the $C_4$-symmetry breaking effect, we add a term $\mathcal{H}_p=\delta S_x$ to the Bloch Hamiltonian in Eq. (\ref{3DBlochHam}), which corresponds to the in-site coupling described by the lattice Hamiltonian $\hat{H}_p=i\delta\sum_{\boldsymbol{r}}\hat{a}^{\dag}_{\boldsymbol{r},\downarrow}\hat{a}_{\boldsymbol{r},0}+\text{H.c.}$. Under the symmetry breaking perturbation, each quadratic DTP splits into two linear triple points. By varying the parameters $M$ and $\delta$, we can obtain the phase diagram of the 3D lattice system, as shown in Fig. \ref{phase-diagram}. In the phase diagram, apart from a normal band insulating phase (denoted by NI) with $C_{k_z}=0$ and a weak topological insulating phase (denoted by TI) with $C_{k_z}=4$ for all range of $k_z$, there are another three different phases: the DTP metal phase (denoted by DTM), the triple-point metal phase with four triple points (denoted by TM$_4$) and triple-point metal phase with eight triple points (denoted by TM$_8$). For example, for $0<\delta<2$ and $1-M<\delta<3-M$, the four triple points are located at $(k_x,k_y,k_z)=(-\arccos(1-\delta),0,\pm\arccos(M-2+\delta))$
and $(\arccos(1-\delta),0,\pm\arccos(M-2+\delta))$. For the TM$_{8}$ phase when $1<\delta<2$, the eight triple points are located at
$(-\arccos(1-\delta),0,\pm\arccos(M-2+\delta))$,
$(\arccos(1-\delta),0,\pm\arccos(M-2+\delta))$,
$(\pi,-\arccos(\delta-1),\pm\arccos(M+2-\delta))$,
and
$(\pi,\arccos(\delta-1),\pm\arccos(M+2-\delta))$.

\begin{figure}[tbph]
\centering
\includegraphics[width=8.5cm]{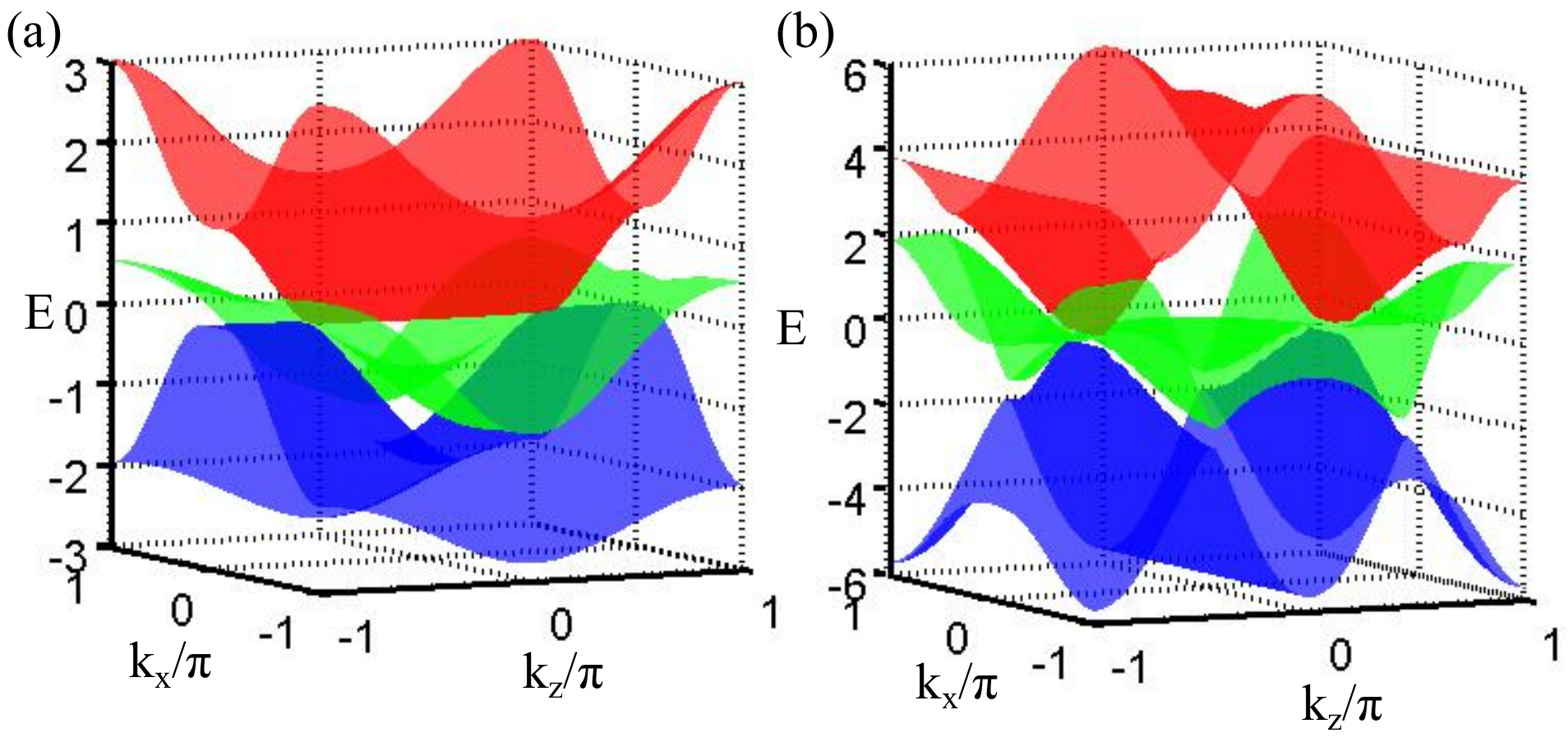}
\caption{(Color online) (a) The energy band in the $k_x$-$k_z$ plane with $k_y=0$: (a) type-II DTPs for $\alpha=0.5$, $\beta=1$ and $M=1.8$; (b) type-III DTPs for $\alpha=1$, $\beta=3$ and $M=1.8$.} \label{typeII}
\end{figure}

As analyzed in Sec. II, we can add an  additional spin-tensor term
to induce the phase transition between type-I and type-II DTPs.
The required spin-tensor term $v_{z}k_{z}N_{zz}$ (let $\beta=1$)
in the continuum Hamiltonian (\ref{DTPHam}) corresponding to the lattice
Hamiltonian
\begin{equation} \label{Nzz}
\begin{split}
\hat{H}_{\mathbf{zz}}&=\sum_{\mathbf{r}}\hat{H}_{\mathbf{rzz}}+\hat{H}_{\mathbf{M_{zz}}},\\
\hat{H}_{\mathbf{rzz}}&=-\frac{1}{2}(\hat{a}^{\dag}_{\mathbf{r},\uparrow}\hat{a}_{\mathbf{r+z},\uparrow}+
\hat{a}^{\dag}_{\mathbf{r},0}\hat{a}_{\mathbf{r+z},0})+\textrm{H.c.},\\
\hat{H}_{\mathbf{M_{zz}}}&=(M-2)\hat{a}^{\dag}_{\mathbf{r},\uparrow}\hat{a}_{\mathbf{r},\uparrow}+
\hat{a}^{\dag}_{\mathbf{r},0}\hat{a}_{\mathbf{r},0}).
\end{split}
\end{equation}
In this case when $1<M<3$ and $0<|\alpha|<1$, there are a pair of
type-II DTPs at $(0,0,\pm k^{c}_{z})$ in the first Brillouin zone,
as shown in Fig. \ref{typeII}(a) with the energy bands. For the 3D
lattice system with the type-II DTPs, we obtain the
$k_z$-dependent Chern number for the lowest band
\begin{eqnarray}\label{typeII1}
&&C_{k_z}=\text{sign}(\alpha)\times\left\{
                                  \begin{array}{ll}
                                    2, ~~& |k_z|>k^{c}_{z};\\
                                    4, ~~& 0<|k_z|<k^{c}_{z}.
                                  \end{array}
                                \right.
\end{eqnarray}
The type-II DTPs with monopole charges $\pm2$ also act as the
transition points between  two layer topological insulators with
Chern number difference $\Delta C_{k_z}=\pm2$.

To induce the transition from  type-I to type-III DTPs, the
required spin-tensor term $\beta v_zk_zN_{xz}$ in Hamiltonian
(\ref{DTPHam}) takes the following form of lattice Hamiltonian
\begin{equation} \label{Nxz}
\begin{split}
\hat{H}_{\mathbf{xz}}&=\sum_{\mathbf{r}}\hat{H}_{\mathbf{rxz}}+\hat{H}_{\mathbf{M_{xz}}},\\
\hat{H}_{\mathbf{rxz}}&=\frac{\beta}{4}(\hat{a}^{\dag}_{\mathbf{r},\downarrow}\hat{a}_{\mathbf{r+z},\uparrow}+
\hat{a}^{\dag}_{\mathbf{r},\downarrow}\hat{a}_{\mathbf{r-z},\uparrow})+\textrm{H.c.},\\
\hat{H}_{\mathbf{M_{xz}}}&=-\frac{\beta(M-2)}{2}(\hat{a}^{\dag}_{\mathbf{r},\downarrow}\hat{a}_{\mathbf{r},\uparrow}+
\hat{a}^{\dag}_{\mathbf{r},\uparrow}\hat{a}_{\mathbf{r},\downarrow}).
\end{split}
\end{equation}
With this addition spin-tensor term, when $1<M<3$ and $|\beta|>2$ ($\alpha=1$), there are a pair of type-III DTPs at $(0,0,\pm k^{c}_{z})$ in the first Brillouin zone, as shown in Fig. \ref{typeII}(b) with the energy bands. For the 3D lattice system with the trivial type-III DTPs, the $k_z$-dependent Chern number $C_{k_z}=0$ for all $k_z$.

\section{Realization and detection in optical lattices}

In this section, we discuss the realization of the lattice  models
and detection of the topological properties of the emergent DTP
fermions in optical lattices. We consider a noninteracting
degenerate fermionic gas in a 3D cubic (or 2D square) optical
lattice, and the three spin states $|\uparrow\rangle$,
$|0\rangle$, $|\downarrow\rangle$ are encoded by three atomic
internal states. By defining the three-component annihilation
operator at site $\mathbf{r}$ as
$\hat{a}_{\mathbf{r}}=(\hat{a}_{\mathbf{r},\uparrow},\hat{a}_{\mathbf{r},0},\hat{a}_{\mathbf{r},\downarrow})^{T}$,
the 3D lattice Hamiltonian in Eq. (\ref{3DHam}) can be rewritten
as
\begin{eqnarray} \label{Ham}
\hat{H}_{3D}&=&\sum_{\boldsymbol{r},\boldsymbol{\eta}}\left(\hat{a}_{\boldsymbol{r}+\boldsymbol{\eta}}^{\dag}U_{\eta}\hat{a}_{\boldsymbol{r}}+\text{H.c.}\right)
+M\sum_{\boldsymbol{r}}\hat{a}_{\boldsymbol{r}}^{\dag}S_z\hat{a}_{\boldsymbol{r}}\\
&&+\sum_{\boldsymbol{r}}\left[\hat{a}_{\boldsymbol{r}+(\boldsymbol{x}+\boldsymbol{y})}^{\dag}U_{xy}\hat{a}_{\boldsymbol{r}}
-\hat{a}_{\boldsymbol{r}+(\boldsymbol{x}-\boldsymbol{y})}^{\dag}U_{xy}\hat{a}_{\boldsymbol{r}}+\text{H.c.}\right]\nonumber,
\end{eqnarray}
where
$\boldsymbol{\eta}=\boldsymbol{x},\boldsymbol{y},\boldsymbol{z}$
denote the hopping directions. The hopping matrices along the
three axis and $xy$ direction are given by
$U_x=\frac{1}{2}(S_x-S_z)$, $U_y=-\frac{1}{2}(S_x+S_z)$,
$U_z=-\frac{1}{2}S_z$, and $U_{xy}=-\frac{1}{4}S_y$, respectively.
For a square lattice without the $U_z$ term, the Hamiltonian
recovers to the 2D model in Eq. (\ref{2DHam}).

The terms $U_{\eta}$ and $U_{xy}$ describe atomic hopping between two lattice sites with spin flipping, which can be achieved by the laser-assisted tunnelling technique with well-designed Raman
coupling between the two spin states \cite{GaugeRMP,GaugeRPP,SOC-Review1,SOC-Review2,Cooper2018}. In experiments, one can first use a moderate magnetic field to distinguish the spin states, and then the natural hopping $t_N$ along each direction is suppressed by titling the cubic optical lattice with a homogeneous energy gradient along the $\eta$ direction, with the large tilt potential $\Delta_{\eta}\gg t_{N}$. The tilt potential can be created through the natural gravitational field or the gradient of a dc- or ac-Stark shift. In order to distinguish the tunnellings directed along
different directions for independent Raman coupling, one requires different linear energy shifts per site $\Delta_{x}\neq\Delta_{y}\neq\Delta_z\neq\Delta_{x}\pm\Delta_{y}$. Finally, the hopping terms can be restored and engineered by application of two-photon Raman coupling with the laser beams of proper configurations through the laser-frequency and polarization selections \cite{GaugeRMP,GaugeRPP,SOC-Review1,SOC-Review2}. The spin-tensor terms in Eqs. (\ref{Nzz}) and (\ref{Nxz}) can also been engineered in this way. In principle, arbitrary hopping matrices including the required hopping terms can be independently created with well-designed laser configurations \cite{GaugeRMP,GaugeRPP,SOC-Review1,SOC-Review2}. Particularly, the detailed Raman-coupling schemes have been proposed to realize similar hopping terms with spin flip $S_{x,y,z}$ for cold atoms in a 3D cubic lattice \cite{Maxwell,HPHu2017,Wang2014}. We note that the realization of all the hopping terms is technically extremely challenging since a considerable number of Raman beams are needed. Although the beams can be drawn from the same laser by an electric or acoustic optical modulator, one should overcome the difficulties in the implementation of the Raman lasers that are associated with the heating for the required spin-changing transitions.

To detect the emergent DTP quasiparticles in the optical lattice, one can first detect the DTP as the band crossings. The 2D and 3D DTPs can be probed via the Bragg spectroscopy or Bloch-Zener oscillations from measuring the atomic zener tunneling to excited band after a Bloch oscillation, similar to the methods used for detecting Dirac points \cite{Zhu2007,Lim,Tarruell2012} and Weyl points \cite{ZDW2015,He,My} in optical lattices. The Berry phase of a 2D DTP can be directly measured using an interferometric approach in momentum space \cite{Duca2015}. Moreover, it has been demonstrated that the full tomography of Bloch states (vectors) can be achieved with cold atoms in optical lattices to reveal the band topology \cite{Alba2011,Hauke2014,Deng2014,Zhang2016,Flaschner2016,Li2016}, which would be applicable in our proposed system.

\begin{figure}[tbph]
\centering
\includegraphics[width=8.5cm]{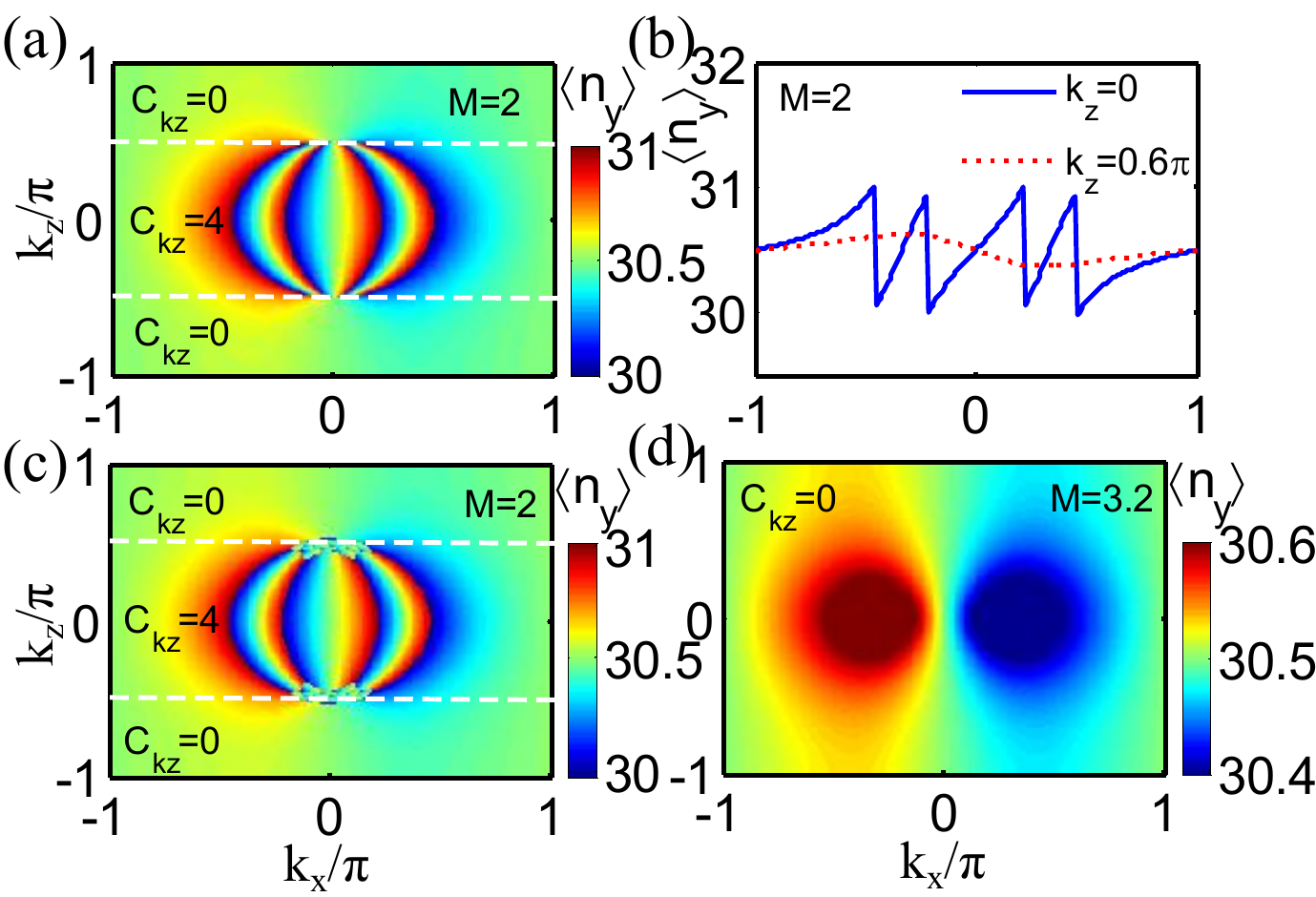}
\caption{(Color online) The hybrid Wannier center $\langle n_{y}(k_x,k_z)\rangle$ in a tight-binding chain of length $L_{y}=60$ under the open boundary condition at $1/3$ filling (the particle number of the three-component fermionic atoms is fixed as $N_a=L_y=60$ in numerical simulations) as a function of the adiabatic pumping parameter $k_{x}$ for different $k_{z}$. (a) The profile $\langle n_{y}(k_x,k_z)\rangle$ for $M=2$ without trapping potential shows four jumps of one-unit-cell for $k_z$ within the region $(-0.5\pi,0.5\pi)$, with typical examples shown in (b). (c) The profile $\langle n_{y}(k_x,k_z)\rangle$ for $M=2$ under a weak harmonic trap with $V_{t}=2\times10^{-4}t$ ($t=1$ as the energy unit). (d) The profile $\langle n_{y}(k_x,k_z)\rangle$ without trapping potential for $M=3.2$ shows the absence of one-unit-cell jumps for all $k_z$ in the topologically trivial regime.} \label{pumping}
\end{figure}

For the 3D case, we show that the $k_z$-dependent Chern number
$C_{k_z}$ can also be measured from the shift of the hybrid
Wannier center of an atomic cloud, based on the particle pumping
approach and hybrid Wannier functions in band theory
\cite{Thouless,Wang,Pumping1,Pumping2,Pumping3,Wanglei,Smith,Marzari}.
For a fixed $k_z$ ($k_z\neq k_z^c$), the Hamiltonian
$H_{k_z}(k_x,k_y)$ describes a 2D insulating lattice system, which
can be viewed as a fictitious 1D insulator along $y$, subject to
the external parameter $k_x$ and $k_z$. The polarization of this
1D insulator can be defined by means of hybrid Wannier functions
\cite{Smith,Marzari,Wanglei}, which are localized in the $y$ axis
retaining Bloch character along $k_x$ and $k_z$. When $k_x$ is
adiabatically changed by $2\pi$, the change in polarization, i.e.,
the shift of the hybrid Wannier center, is proportional to the
$k_z$-dependent Chern number, which is a manifestation of
topological pumping with $k_x$ being the adiabatic pumping
parameter. Considering the Bloch Hamiltonian in Eq.
(\ref{3DBlochHam}) with parameters $k_x$ and $k_z$, and
transforming it to the tight-binding Hamiltonian along the $y$
axis, we can construct the hybrid Wannier center as
\cite{Wanglei,My}
\begin{eqnarray}\label{center}
\langle n_y(k_x,k_z)\rangle=\frac{\sum_{i_y}i_y\rho_{i_y}(k_x,k_z)}{\sum_{i_y}\rho_{i_y}(k_x,k_z)},
\end{eqnarray}
where $\rho_{i_y}(k_x,k_z)$ is the density of the hybrid Wannier function as a function of $k_x$ and $k_z$, with $i_y$ being the lattice-site index in the 1D tight-binding chain. Here, the hybrid density can be written as
\begin{eqnarray}\label{density}
\rho_{i_y}(k_x,k_z)=\sum_{\text{occ}}|k_x,k_z\rangle_{i_yi_y}\langle k_x,k_z|,
\end{eqnarray}
where $|k_{x},k_{z}\rangle_{i_{y}}$ denotes the hybrid wave
function of the system at site $i_y$ and notation occ denotes the
occupied states. Experimentally, the atomic density
$\rho_{i_y}(k_x,k_z)$ can be directly detected by the hybrid
time-of-flight images \cite{Wanglei}, combing \emph{in situ}
imaging along $y$ and time-of-flight imaging along the release
directions $x$ and $z$. In the measurement, the optical lattice is
switched off along $x$ and $z$, while the system remains unchanged
along $y$. Thus, one can directly extract $C_{k_z}$ from an
experimental detection of the shift in the hybrid function center.

We numerically calculate $\langle n_y(k_x,k_z)\rangle$ from Eq. (\ref{center}) in a 1D reduced
tight-binding chain of length $L_y=60$ at $1/3$ filling (the lowest band is filled), with
typical results shown in Fig. \ref{pumping}. Here we note that for three-component fermionic atoms in the three-band model with the total lattice site $N_L$, the particle number is $N_a=N_L$ (the average atomic density is $1$) under the 1/3 filling condition. In our numerical simulation for the reduced 1D chain with parameters $k_x$ and $k_z$, the fermion number is thus fixed as $N_a=L_y=60$ for various $k_x$ and $k_z$. To consider the effect of the fluctuations of $\rho_{i_y}(k_x,k_z)$ in realistic experiments, we simply add local fluctuations as $(1+w_{i_y})\rho_{i_y}(k_x,k_z)$ in our numerical calculations, where the random parameter $w_{i_y}$ for the cite $i_y$ is uniformly distributed in the range $[-W,W]$ with the fluctuation strength $W$. We have numerically checked that the results of $\langle n_y(k_x,k_z)\rangle$ shown in Fig. \ref{pumping} can preserve when the fluctuation strength $W\lesssim30\%$ ($W\lesssim5\%$) for the quantity averaged over hundreds of random samplings (a single random sampling). Thus, such a magnitude of the density fluctuations would not affect the measurements in practical cold-atom experiments after averaging the samples.

For the case of $M=2$ in Figs. \ref{pumping}(a) and (b), the results show that the system exhibits four discontinuous
jumps of one unit cell within the region
$k_z\epsilon(-k_z^c,k_z^c)$ with $k_z^c=0.5\pi$, as $k_x$ changing
from $-\pi$ to $\pi$. This indicates that when
$k_z\epsilon(-k_z^c,k_z^c)$, $C_{k_z}=4$ and outside this region
$C_{k_z}=0$. For $M=3.2$ in
Fig. \ref{pumping}(d), there is no discontinuous jump in $\langle n_y(k_x,k_z)\rangle$,
which indicates that the system is in the trivial insulator phase with
$C_{k_z}=0$ for all $k_z$. To take the realistic experiment into
account, we add a weak harmonic trap
$\hat{H}_t=V_t\sum_{n}(n-\frac{L_y}{2})^{2}\hat{a}^{\dag}_n\hat{a}_n$
to this finite-site lattice, where $V_t$ is the trap strength.
Within a local-density approximation, the lowest band is still
filled at the center of the trap and thus the shift of the hybrid
Wannier center is expected to be nearly the same as those without
the trap potential. When the band gap $E_g\leq
V_t(i_y-\frac{L_y}{2})^{2}$, the lowest band is only partially
filled near the two edges and then this pumping argument does not
apply to this region. With numerical simulations shown in Fig.
\ref{pumping}(c), $\langle n_y(k_x,k_z)\rangle$ preserve with a
deviation less than $2\%$ except the regions near the DTPs for
$V_t=2\times10^{-4}t$, which are consistent with the estimations in
the local-density analysis.

\section{Conclusions}

In summary, we have proposed a class of pseudospin-1  quadratic
DTP fermions emerging in topological metal bands. We have analyzed
a general three-band continuum model with $C_4$ symmetry in three
dimensions, which has three types of threefold DTPs as spin-1
generalization of double-Weyl points, classified by their
topological charges. The 2D and 3D tight-binding lattice models of
topological metal bands with exotic DTP fermions near the DTPs
have also been proposed and explored. In 2D, the bands close at a
trivial DTP with zero Berry phase, which occurs at the transition
between the normal and topological insulator phases. In 3D, the
topological properties of three different DTP fermions in lattice
systems are further investigated, and the effects of breaking
$C_4$ symmetry are also considered, which generally leads to
splitting each quadratic DTP into two linear triple points and
gives topological phase diagrams. Finally, we have discussed
realization of the proposed models and detection of the
topological properties of the DTP fermions in optical lattices.

\acknowledgements{This work was supported by the NSFC (Grant No.
11604103, No. 11704132, No. 11474153, and No. 91636218), the NKRDP
of China (Grant No. 2016YFA0301803), the NSAF (Grant No. U1830111), the NSF of Guangdong Province
(Grant No. 2016A030313436), the Startup Foundation of SCNU, and
the Innovation project of Graduate School of SCNU.}

\end{document}